\documentclass[11pt,nofootinbib,aps,prd]{revtex4}
\usepackage{amsmath,amssymb}
\usepackage[mathcal,mathscr]{eucal}
\usepackage{latexsym}

\begin{document}
\thispagestyle{empty}
\vspace*{-10mm}
\begin{flushright}
YITP-09-37
\end{flushright}

\newcommand{\todo}[1]{{\em \small {#1}}\marginpar{$\Longleftarrow$}}
\newcommand{\labell}[1]{\label{#1}\qquad_{#1}} 

\vskip 1cm

\begin{center}
{\Large \bf Tensor ghosts in the inflationary cosmology}
\end{center}
\vskip 1cm

\renewcommand{\thefootnote}{\fnsymbol{footnote}}
\centerline{\bf
 Tim Clunan\footnote{T.P.Clunan{}@{}damtp.cam.ac.uk}
 and Misao Sasaki\footnote{Misao{}@{}yukawa.kyoto-u.ac.jp}}
\vskip .5cm
\centerline{ \it Department of Applied Mathematics and Theoretical Physics}
\centerline{ \it Cambridge University, Cambridge, CB3 0WA, UK}
\vskip .5cm
\centerline{ \it Yukawa Institute for Theoretical Physics}
\centerline{\it Kyoto University, Kyoto 606-8502, Japan}

\setcounter{footnote}{0}
\renewcommand{\thefootnote}{\arabic{footnote}}

\begin{center}
\bf Abstract
\end{center}
Theories with curvature squared terms in the action
are known to contain ghost modes in general. 
However, if we regard curvature squared
terms as quantum corrections to the original theory, the emergence
of ghosts may be simply due to the perturbative truncation of a
full non-perturbative theory. If this is the case, there should be
a way to live with ghosts. In this paper, we take the Euclidean path 
integral approach, in which ghost degrees of freedom can be, 
and are integrated out in the Euclideanized spacetime.
We apply this procedure to Einstein gravity with a Weyl curvature 
squared correction in the inflationary background.
We find that the amplitude of tensor perturbations is
modified by a term of $O(\alpha^2H^2)$ where $\alpha^2$ is
a coupling constant in front of the Weyl squared term and $H$ is
the Hubble parameter during inflation.

\vspace{5mm}
\section{Introduction}

Previously in the literature corrections to the Einstein-Hilbert action formed
with terms of higher order in the Riemann tensor have been considered. 
In the past these have been considered in connection with the possibility of 
a non-local theory \cite{Simon:1990ic} and also 
renormalisation \cite{DeWitt:1980hx, Dowker:1975tf, Hawking:1976ja}. These
higher derivative  theories do, however, suffer from the problem of ghosts. 
There are, of course, `good' ghosts, such as the Faddeev-Popov ghost; but 
higher derivative ghosts are `bad' ghosts. Even if one adds `small' higher 
derivative terms, so that the lagrangian is almost the same as the second 
order theory, the resulting theory is not well described perturbatively since
there are now extra degrees of freedom. These bring with them a number of 
problems \cite{breakdown, Hawking:2001yt, Hawking:1985gh, Simon:1990ic}. 
They are negative energy particles, with energies which blow-up in the limit 
where the higher derivative terms are removed from the Lagrangian. 
Ghosts also form states of negative norm and hence lead to non-unitarity. 

The simple harmonic oscillator can be made to mimic the effect of these terms 
by adding higher derivative terms to its Lagrangian. This was considered by 
Hawking and Hertog \cite{Hawking:2001yt}; they showed that despite these 
problems it is possible to obtain a sensible probability distribution for 
observations of the field. This result relied on the Euclidean path integral 
formulation of quantum theory. This, of course, echoes the No-boundary 
cosmology of Hartle and Hawking~\cite{Hartle:83}. Other authors have considered the effect of extra terms in the lagrangian. 
In the cosmological case the full theory is not taken seriously, and instead 
back-substitution \cite{Weinberg:08,breakdown} is typically used. Here we 
run contrary to this trend by extending and comparing the result of Hawking 
and Hertog to the cosmological case where the Lagrangian contains 
a Weyl tensor squared term.

It is normally the case for ghosts that even if one could deal with the 
non-interacting theory, the interacting theory would have the problem of
run-away production of positive energy and negative energy particles. 
This would require further work, but is not expected to be a problem in the Euclidean approach where a late time boundary condition is specified \cite{Hawking:2001yt, Coleman:1969xz}.

The very early Universe provides us with the ultimate high energy physics 
experiment. Post inflation the energies are sufficiently low that one could
not rule out the occurence of higher order curvature terms in 
the Lagragian; however, during inflation these terms could be of 
importance. One could consider a whole series of terms of arbitrary order
in the Riemann tensor, but it makes sense to start with just second order.

Having accepted that there is the possibility of higher curvature terms, 
the question is how one should deal with them. One can handle them simply 
by back-substituting the modes from the usual equations of motion. 
This approach has been used in \cite{breakdown}. Here one finds a 
momentum cut-off much higher than the Plank energy.

On the other hand \cite{flannagan_wald} argue that by using a perturbative 
approach, rather than solving the full system, one can miss essential 
physical features. For example, in the case of the heat equation initial 
data of compact support lead to a perturbative solution with the same 
support; however, the actual solution clearly spreads.

There is something special about the case where the only additional terms we consider are curvature squared terms. This theory has better renormalisation properties \cite{Stelle:1976gc}.  We deal with the slow roll case in four dimensions and therefore restrict our attention to the Weyl tensor squared terms.

The structure of this paper is as follows. In \S\ref{sec:hdsho} we review the simple harmonic oscillator. In \S\ref{sec:prelim} we find the action. In \S\ref{sec:eom} we solve the eom. In \S\ref{sec:The_wavefunction_for_a_de_Sitter_background} we find the wavefunction. In \S\ref{sec:tensor_observables} we find the two point function; and in \S\ref{sec:conc} we draw our conclusions.

\section{The higher derivative simple harmonic oscillator}
\label{sec:hdsho}

In \cite{Hawking:2001yt} the authors consider a higher derivative simple harmonic oscillator, the extra terms being set up to emulate the situation in gravity. This has the advantage of fewer terms in the lagrangian and doing away with issues to do with gauge, thus providing clarity in the exposition of the method. One can expect that the method works in the case of gravity, and if one is not interested in the detailed form of the resulting observables it is more than adequet. However, as future experiments advance one may expect to be more interested in these details, which an ersatz theory cannot provide. Wishing to examine the form of the result more closely in gravity, the natural place to start is the tensor sector as it is already gauge invariant.

Hawking and Hertog find that despite the problems presented by typical higher derivative theories it is possible to take the full theory seriously (not removing any modes) and arrive at a sensible probability distribution. Key to this method is the fact that they are considering the field as a Euclidean field.

The action for a Euclidean higher derivative simple harmonic oscillator is
\begin{equation}
I=\int d \tau \, \left(\frac{\alpha^2}{2} \phi_{,\tau\tau}^2 +\frac{1}{2} \phi_{,\tau}^2 + \frac{1}{2} m^2 \phi^2 \right) \, ;
\end{equation}
and, although this theory doesn't contain any interaction terms the authors argue that these would not cause further problems in the case where we are prescribing the field value on a late time surface. For real $\phi$ and real $\tau$ this action is clearly positive semi-definite and hence results in a convergent path integral. The resulting equation of motion
\begin{equation}
\alpha^2 \phi_{,\tau\tau\tau\tau} - \phi_{, \tau\tau} + m^2 \phi = 0
\end{equation}
has a general solution which is easily seen to be
\begin{equation}
\phi (\tau)
 = A \, \sinh (\lambda_+ \tau) + B \, \cosh (\lambda_+ \tau) 
+ C \, \sinh (\lambda_- \tau) + D \, \cosh (\lambda_- \tau)
\end{equation}
where
\begin{equation}
\lambda_{\pm} = \frac{1}{\sqrt{2 \alpha^2}}
 \sqrt{1 \mp \sqrt{1 - 4 m^2 \alpha^2}} \approx
\left\{\begin{array}{l}
m\,\\
 \alpha^{-1}
\end{array}\right.
\end{equation}
for small alpha (which is our area of interest since this 
results in a theory which has `small' higher derivative terms). 
The ground state wave function at $\tau=0$, $\Psi_{\tau =0}$,
 is the amplitude to go from zero $\phi (\tau = -\infty) =0$, 
$\phi_{, \tau} (\tau = -\infty) =0$ to prescribed values 
$\phi_0$, $\phi_{0 , \tau}$ at $\tau = 0$. 
The authors calculate this using the stationary phase
 approximation (and are thus assesing on a classical solution), 
\begin{equation}
\Psi_0 (\phi_0 , \phi_{0, \tau}) 
= N \exp \left(-F \left( \phi_{0, \tau}^{2}
 + \frac{m}{\alpha} \phi_{0}^2 \right)
 + \frac{2 m^2 - m/\alpha}{(\lambda_{-} - \lambda_{+})^2}
 \phi_0 \phi_{0,\tau} \right)
\end{equation}
where 
\begin{equation}
F = 
\frac{1- 4 m^2 \alpha^2}{2 \alpha^2 (\lambda_{+} +\lambda_{-})
(\lambda_{-}-\lambda_{+})^2} \, .
\end{equation}
The Euclidean conjugate ground state wave function $\Psi^*$ is
 defined in a similar way, but integration goes from 
$\tau = 0$ to $\infty$; it is the analogue of complex conjugating
 the Lorentzian wavefunction. The associated probability is thus 
\begin{equation}
P(\phi_0 , \phi_{0, \tau}) = \Psi \Psi^* 
= N^2 \exp \left( -2 F \left( \phi_{0 , \tau}^2
 +\frac{m}{\alpha} \phi_0^2 \right) \right) \, , 
\end{equation}
and we see that we may integrate over the unobserved 
$\phi_{0 , \tau}$. If we had been working with $\phi_{0 , t}$,
 where $t$ is the Lorentzian time, the sign infront of the 
$\phi_{0 , t}^2$ term would have been wrong and integration 
not possible: the possibility is afforded by working in the 
Euclidean formalism. This results in a probability
\begin{align}
P(\phi_0) &= \sqrt{\frac{2 F m}{\pi\alpha}} 
\exp \left( -\frac{2 m F}{\alpha} \phi_0^2 \right)\\
& \approx \sqrt{\frac{m}{\pi}}
 \left(1+\frac{m\alpha}{2}\right) 
\exp (-m(1+m\alpha) \phi_0^2) \, ,
\end{align}
the approximation holding for small $\alpha$. So we see that 
the second order theory is corrected by terms in $\alpha$, 
rather than $\alpha^2$ as we would see had we back-substituted
 the modes from the second order theory into the fourth
 order theory.

\section{Preliminaries}
\label{sec:prelim}
We are interested in actions with terms up to second order in the
curvature tensor. Since we are working in four dimensions the
Gauss-Bonnet term is a total derivative and so we may replace 
$R_{\mu\nu}R^{\mu \nu}$ with a combination of $R^2$ and 
$C^{\alpha \beta \mu\nu} C_{\alpha \beta \mu \nu}$. 
The $R^2$ term leads to an additional massive scalar field 
which doesn't cause any problems.\footnote{One might 
expect a problem with the $R^2$ term; however, the only metric
 component having a non-degenerate higher derivative term due 
to this is the gravitational potential energy in the usual theory.
The extra modes due to the higher derivatives therefore have
positive energy.}
 We are interested in the negative energy ghost component, 
which Boulware \cite{Boulware:83}
has shown to be contained in the tensor part of the Weyl 
squared term. Thus, we focus on this. 

The York-Gibbons-Hawking boundary term $B_{YGH}$ is usually 
added to the Einstein-Hilbert action in order to make the 
variational problem (where one chooses the metric on a 
spacelike surface) well posed \cite{York:72,Gibbons:77}. 
This comes down to the fact that if 
we are specifying `$q$' at initial and final times we 
do not also wish to specify `$\dot{q}.$' 
In the context of the higher derivative theory we 
condsider here we will specify `$q$' and `$\dot{q},$' 
so at first glance it might seem that we do not need
 $B_{YGH}$; however, since this term 
also makes the Euclidean action for the inhomogeneous 
modes positive definite \cite{Hawking:2000ee} it 
transpires that we will need it.

At zeroth order in slow roll the spacetime during inflation
reduces to de Sitter space. We use the metric 
\begin{equation}
\label{metric}
ds^2 =e^{2 \rho}( -d\eta^2 + (\delta_{i j}+\gamma_{i j})dx^{i}dx^{j})\,,
\end{equation}
where $\gamma_{i j}$ is a transverse traceless perturbation and we can
conveniently take $e^{2\rho}$ to be $1$ or $1/(H\eta)^2$ according to 
whether we wish to discuss flat or de Sitter space. From 
Boulware \cite{Boulware:83} we have 
\begin{equation}
C^{\alpha \beta \mu \nu} C_{\alpha \beta \mu \nu}=8 C^{0 k 0 l}
C_{0 k 0 l}+4 C^{0 k l m} C_{0 k l m} \, .
\end{equation}
This leads to 
\begin{align}
C^{\alpha \beta \mu \nu} C_{\alpha \beta \mu \nu}
&=
{1\over 2}
e^{-4\rho} \big((\gamma''_{i j}+\gamma_{i j ,n n})
(\gamma''_{ij}+\gamma_{i j ,m m} )-2(\gamma'_{i j,k}-\gamma'_{i k,j})
(\gamma'_{ij,k}-\gamma'_{i k,j} )\big)\\
\label{csq2}
&={1\over 2}e^{-4\rho}\big(\gamma''_{i j}\gamma''_{i j}
+2\gamma_{i j,nn}\gamma''_{i j}+4 \gamma'_{i j}\gamma'_{i j, k k}
+\gamma_{i j,n n}\gamma_{i j,m m}
\big)\,,
\end{align}
at second order, where spatial boundary terms have been
 dropped.\footnote{We are only ever concerned with the fields
on future and past boundaries of a spacetime volume in the 
definition and propagation of the wavefunction.}
Here and in what follows the repeated spatial indices 
$i,j,k,\ldots$ are to be summed over.
Since the metric is conformally flat
the Weyl tensor is vanishing at zeroth order and hence it is only
necessary to calculate it to first order if one is finding the $C^2$
term to second order. 

Considerations so far lead us to start with the action
\begin{equation}
\label{initial_action}
S=\int\sqrt{g}\,d^4 x \Big({1\over 2}R-\Lambda
- \alpha^2 C^{\alpha \beta \mu \nu} C_{\alpha \beta \mu \nu}\Big)+B_{YGH}\,.
\end{equation}
where 
\begin{equation}
B_{YGH} = \int d^3 x \sqrt{h} K \, ;
\end{equation}
however, we will need an additional boundary term relating
to the $C^2$ term.

\subsection{Action for flat space}
If $\Lambda=0$, the Minkowski spacetime is a solution of the theory.
On this background the action (\ref{initial_action}) becomes 
\begin{eqnarray}
&&S=
\int d \eta d^3 x \Biggl[{1\over 8}
(\gamma'_{i j}\gamma'_{ij}-\gamma_{i j,k}\gamma_{i j,k})
\cr
&&\qquad\qquad
-{\alpha^2\over 2}\Bigg(\gamma''_{i j}\gamma''_{i j}
+2\gamma_{i j,nn}\gamma''_{i j}-4 \gamma'_{i j,k}\gamma'_{i j, k}
+\gamma_{i j,n n}\gamma_{i j,m m}\Bigg)\Biggr]
\end{eqnarray}
at second order in the gravitational wave perturbation. We have dropped
spatial boundary terms as these vanish for localised fields. We only
consider the transverse traceless perturbations, and at second order
these decouple from the scalar and vector perturbations. This gives us
the ghost content of the theory. Considering now $\imath S$ (the argument 
of the path integral defining the wavefunction) rotated into the Euclidean
with $\tau = \imath \eta$ being the Euclidean `time' we see that the 
only term preventing the Euclidean action $I=-\imath S$ 
being positive definite on real Euclidean fields is 
\begin{equation}
-\int d \eta d^3 x \,\alpha^2 \gamma_{i j,n n}\gamma''_{i j}\,.
\end{equation}
This leads us to the addition of a boundary term on the initial and 
final boundary surfaces. In this way we can do an integration by parts 
in the time coordinate and have a positive definite Euclidean 
action, so the path integral will be well defined. This boundary term is 
the same in the de Sitter case.

\subsection{Action for de Sitter space}
The action in the case of the metric (\ref{metric}) is 
\begin{align}
\nonumber
S=&\int d\eta d^3 x \Biggl[
-3 e^{2\rho} \rho'^2 -e^{4 \rho} \Lambda\\
\label{de_sitter_action}
&\;\;\;+{1\over 8} e^{2 \rho}
\Big(\gamma'_{i j}\gamma'_{i j}-\gamma_{i j,k}\gamma_{i j,k}\Big)
 - {\alpha^2 \over 2} \Big(\gamma''_{i j}\gamma''_{i j}+
2 \gamma'_{i j}\gamma'_{i j, k k} +\gamma_{i j,n n}\gamma_{i j,m m} \Big)
\Biggr]\,,
\end{align}
where the background equation of motion (eom) holds. 
The eom for the tensor fluctuations is thus
\begin{equation}
\frac{1}{4}\left((e^{2\rho}\gamma_{i j}')'-e^{2\rho}\gamma_{i j , k k}\right)
+\alpha^2 \left(\gamma_{i j}''''-2\gamma_{i j , k k}''
+\gamma_{i j , m m n n}\right) =0\,.
\end{equation}
So expanding $\gamma$ as 
\begin{equation}
\label{fourier}
\gamma_{i j} =\int {d^3 k \over (2\pi)^3} \sum_{s=\pm}
\epsilon^{s}_{i j}(\vec{k}) \gamma^{s}_{\vec{k}}(\eta) e^{\imath \vec{k}\cdot
\vec{x}}\,,
\end{equation}
where $\epsilon^{s}_{i i}(\vec{k})=0=k^{i}\epsilon^{s}_{i j}(\vec{k})$,
 $\epsilon^{s}_{i j}(-\vec{k})=\epsilon^{s\; *}_{i j}(\vec{k})$
 and 
$\epsilon^{s}_{i j}(\vec{k})\epsilon^{t\; *}_{i j}(\vec{k})=2 \delta_{s t}$ 
we find 
\begin{equation}
\label{eom_fourier}
\frac{1}{4}\left((e^{2\rho}\gamma_{\vec{k}}^{s \, '})'
+k^2 e^{2\rho}\gamma_{\vec{k}}^s\right)
+\alpha^2 \left(\gamma_{\vec{k}}^{s\,''''}+2 k^2\gamma_{\vec{k}}^{s\,''}
+k^4 \gamma_{\vec{k}}^s\right) =0\,.
\end{equation}

Classically $\gamma_{i j}$ is taken as a real field. Then the 
second order action may be written in terms of the fourier components as 
\begin{equation}
\label{2nd_order_ds_fourier_action}
S_{\vec{k}}=\sum_{s=\pm} \int d \eta 
\Bigg[{1\over 4} e^{2 \rho}
(|\gamma_{\vec{k}}^{s\,'}|^2-k^2 |\gamma_{\vec{k}}^{s}|^2)
-\alpha^2 (|\gamma_{\vec{k}}^{s\,''}|^2-2 k^2 |\gamma_{\vec{k}}^{s\,'}|^2
+k^4 |\gamma_{\vec{k}}^{s}|^2)\Bigg],
\end{equation}
which we regard as an action for the real and imaginary parts
of $\gamma_{\vec{k}}^s$ as independent fields. 

\subsection{Canonical formalism}
In \cite{Hawking:2001yt} it is argued that whilst most authors considering
the canonical formalism for fourth order theories would take `$q$' 
and `$\ddot{q}$' to be the canonical `position' coordinates, in the path
integral formalism one should describe a state by `$q$' and `$\dot{q}$' 
at initial and final times. With this in mind we consider
$Q_{\gamma_{\vec{k}\,r}^s}=\gamma_{\vec{k}\,r}^s$ and 
$Q_{\gamma_{\vec{k}\,r}^{s\,'}}=\gamma_{\vec{k}\,r}^{s\,'}$; 
these have conjugate momenta
\begin{align}
P_{\gamma_{\vec{k}\,r}^s}&
=2\left({1\over 4}e^{2\rho} \gamma_{\vec{k}\,r}^{s\,'}
 + \alpha^2 \gamma_{\vec{k}\,r}^{s\,'''}
 + 2\alpha^2 k^2 \gamma_{\vec{k}\,r}^{s\,'}\right),
\\
P_{\gamma_{\vec{k}\,r}^{s\,'}}&=-2\alpha^2 \gamma_{\vec{k}\,r}^{s\,''},
\end{align}
and similarly for the imaginary components of the fields. 
Thus we have a Hamiltonian
\begin{eqnarray}
\mathscr{H}_{\vec{k}}^s 
&=& - \alpha^2 |\gamma_{\vec{k}}^{s\,''}|^2 
+2\alpha^2 \Re(\gamma_{\vec{k}}^{s\,'''\,*} \gamma_{\vec{k}}^{s\,'})
+{e^{2\rho}\over 4}  |\gamma_{\vec{k}}^{s\,'}|^2 
+2 \alpha^2 k^2 |\gamma_{\vec{k}}^{s\,'}|^2 
+{e^{2\rho} k^2 \over 4}  |\gamma_{\vec{k}}^s|^2
 +\alpha^2 k^4 |\gamma_{\vec{k}}^s|^2
\label{classical_hamiltonian_gamma} 
\\
&=&
\Re (P_{\gamma_{\vec{k}}^s}^* Q_{\gamma_{\vec{k}}^{s\,'}})
-{1 \over 4 \alpha^2} |P_{\gamma_{\vec{k}}^{s\,'}}|^2
-{e^{2 \rho}\over 4}  |Q_{\gamma_{\vec{k}}^{s\,'}}|^2
+{e^{2\rho} k^2\over 4}  |Q_{\gamma_{\vec{k}}^s}|^2
-2\alpha^2 k^2 |Q_{\gamma_{\vec{k}}^{s\,'}}|^2 
+ \alpha^2 k^4 |Q_{\gamma_{\vec{k}}^s}|^2 ,
\label{classical_hamiltonian} 
\end{eqnarray}
where again we take the independent variables to be the real 
and imaginary components of each field.

Taking the classical Hamiltonian (\ref{classical_hamiltonian})
 one sees the ghost instability is
present because the only occurence of $P_{\gamma}$ is in the
$P_{\gamma}Q_{\gamma'}$ term and hence the Hamiltonian can be made
arbitrarily negative by fixing $Q_{\gamma'}\ne 0$ and taking
$P_{\gamma}$ appropriately large positive or negative. Unlike the case
of a particle orbiting in a central potential,\footnote{Of course, 
it is also the case that the classical hydrogen atom Hamiltonian can 
be made arbitrarily negative by taking the electron to have small 
momentum and be close enough to the nucleus. This problem with a small 
region of phase space suggests that the hydrogen atom has states of
 arbitrarily negative energy. Heisenberg's uncertainty principle stops 
a state being localised on this problematic region, and so there is a 
ground state.} 
this problem is present for a large volume of phase space, and it is 
this which makes the difference \cite{Woodard:06}. 
Quantizing the higher derivative theory will not result in a lower 
bound on the energy of states.


\subsection{Flat space wavefunction}
In flat space the eom (\ref{eom_fourier}) becomes
\begin{equation}
0={1\over 4}\gamma_{\vec{k}}^{s\,''}+{1\over 4} k^2 \gamma_{\vec{k}}^s
 +\alpha^2 \Big(\gamma_{\vec{k}}^{s\,''''}+2 k^2 \gamma_{\vec{k}}^{s\,''}
+ k^4 \gamma_{\vec{k}}^s \Big),
\end{equation}
which is easy to solve in terms of exponentials. It factorizes to 
give solutions with $k_{+} = k$ and $k_{-}=\sqrt{k^2+1/(4 \alpha^2)}$.

The Lorentzian flat space wavefunction for the mode $\gamma_{\vec{k}}^{s}$ is 
\begin{align}
\nonumber
\Psi_{\vec{k}}^{s}(Q_{\gamma_{\vec{k}}^s},Q_{\gamma_{\vec{k}}^{s\,'}})
=&
N(\eta) \exp\Bigg( -\alpha^2 k_{+}k_{-}
(k_{+}+k_{-})|Q_{\gamma_{\vec{k}}^s}|^{2}
\\
&-\imath \alpha^2 k_{-}k_{+}(Q_{\gamma_{\vec{k}}^s}^* 
Q_{\gamma_{\vec{k}}^{s\,'}}+Q_{\gamma_{\vec{k}}^s}
Q_{\gamma_{\vec{k}}^{s\,'}}^* )
+ \alpha^2 (k_{+}+k_{-})|Q_{\gamma_{\vec{k}}^{s\,'}}|^{2} \Bigg),
\end{align}
which may be obtained through the Euclidean path integral 
prescription, followed by rotation back to Minkowski time, and
 satisfies the Wheeler-de-Witt equation,
\begin{equation}
\mathscr{H}_{\vec{k}}^{s} \Psi_{\vec{k}}^{s}
 = \imath \partial_{\eta} \Psi_{\vec{k}}^{s}
\end{equation}
with
\begin{eqnarray}
\label{flat_hamiltonian}
&&\mathscr{H}_{\vec{k}}^s
 =\Re (P_{\gamma_{\vec{k}}^s}^* Q_{\gamma_{\vec{k}}^{s\,'}})
-{1 \over 4 \alpha^2} |P_{\gamma_{\vec{k}}^{s\,'}}|^2
-{1\over 4}  |Q_{\gamma_{\vec{k}}^{s\,'}}|^2
\cr
&&\qquad\qquad
+{k^2\over 4}  |Q_{\gamma_{\vec{k}}^s}|^2
-2\alpha^2 k^2 |Q_{\gamma_{\vec{k}}^{s\,'}}|^2 
+ \alpha^2 k^4 |Q_{\gamma_{\vec{k}}^s}|^2 ,
\end{eqnarray}
where, as usual, $P = - \imath \partial_{Q}$. With the boundary term 
chosen to make the path integral defining the wavefunction 
well defined we will see that this is also true in the de-Sitter case.

\section{Solving the equation in the de-Sitter case.}
\label{sec:eom}

The equation of motion (\ref{eom_fourier}) for $\gamma_{\vec{k}}^s$ with
$e^{2\rho}=\frac{1}{H^2 \eta^2}$ has the two convenient factorizations,
\begin{align}
\label{eom_fact_1}
&0=\Big({d^2\over d z^2}+{2\over z}{d \over d z}+1-{1\over 4 \beta z^2}\Big)
\Big({d^2\over d z^2} - {2\over z}{d \over d z} +1\Big)
\gamma_{\vec{k}}^s\,,
\\
\label{eom_fact_2}
&0=\Big({1\over z^2}{d^2\over d z^2} - {2\over z^3}{d \over d z}
+{1\over z^2}\Big)
\Big(z^2 {d^2\over d z^2}-2 z{d \over d z}+2+z^2 -{1\over 4 \beta}\Big)
\gamma_{\vec{k}}^s\,,
\end{align}
where $z=-k \eta$ and $\beta = - H^2 \alpha^2$. Hence the solutions
vanishing in the upper half $\eta$ plane are
\begin{align}
\label{mode_solutions_1}
&(1+\imath z) e^{-\imath z}\,,
\\
\label{mode_solutions_2}
&z^{3/2} \Big(J_{{1\over 2}\sqrt{1+1/\beta}}(z)
 -\imath Y_{{1\over 2}\sqrt{1+1/\beta}}(z)  \Big):
\end{align}
these each also solve the first factor in each of (\ref{eom_fact_1}),(\ref{eom_fact_2}).
Whilst the mode (\ref{mode_solutions_1}) is conveniently the same as in 
the Einstein-Hilbert case, we will see that it differs in normalisation. One can also see that this mode is an eigenfunction of the first factor of  (\ref{eom_fact_2}) with non-zero eigenvalue and thus obviously solves (\ref{eom_fact_2}). 
One can see from the second order equations these satisfy that they both
also obay an equation of the form,
\begin{equation}
\frac{1}{z^2}\Big(f^{*}\frac{d}{dz} f - f \frac{d}{dz} f^{*} \Big)
=\textrm{const.}.
\end{equation}
This, and similar equations, are useful in the normalization of the modes.

\subsection{Normalizing the modes}

With the canonical `position' coordinates $ Q_{\gamma_{i j}} $
 and $ Q_{\gamma_{i j}'} $ the conjugate momenta are
\begin{align}
P_{\gamma_{i j}}&=  {1\over 4 H^2 \eta^2} \gamma_{i j}'
+\alpha^2 \gamma_{i j}''' - 2\alpha^2 \gamma_{i j , k k}'\,,
\\
P_{\gamma_{i j}'}&=-\alpha^2 \gamma_{i j}''\,.
\end{align}
The position space Hamiltonian these give rise to correctly generates
the canonical Hamiltonian evolution equations, and thus these put the
Poisson brackets in canonical form.
The operator versions of these satisfy the equal time commutation relations
\begin{align}
[ Q_{\gamma_{i j}} (\eta,\vec{x}), P_{\gamma_{i j}}(\eta,\vec{y})] 
& = 2 \imath \delta^{(3)} (\vec{x}-\vec{y})\,,
\\
[ Q_{\gamma_{i j}^{'}} (\eta,\vec{x}), P_{\gamma_{i j}^{'}} (\eta,\vec{y})] 
&= 2 \imath \delta^{(3)} (\vec{x}-\vec{y})\,,
\end{align}
where the factor of $2$ occurs because summing over $i$, $j$ sums over 
the ``plus" and ``cross" modes.  If we use 
\begin{equation}
\gamma_{\vec{k}}^s=u_k a_{\vec{k}}^{s \, \dagger} + u_k^* a_{-\vec{k}}^s
+v_k b_{\vec{k}}^{s\,\dagger} + v_k^* b_{-\vec{k}}^s
\end{equation}
in (\ref{fourier}) the normalized modes,
\begin{align}
u_{k}&={H \over \sqrt{k^3 (1-8\beta)}} (1-\imath k \eta) e^{\imath k \eta}\,,
\\
\label{normalised_mode_v}
v_{k}&={H \over \sqrt{k^3 (1-8\beta)}} \sqrt{{\pi\over 2}}
 e^{{-\imath \pi \over 4}\sqrt{1+{1\over \beta}}} (-k \eta)^{{3\over 2}}
 H_{{1\over 2}\sqrt{1+{1\over \beta}}}^{(2)} (-k \eta)\,,
\end{align}
result in 
\begin{align}
[a_{\vec{k}}^s,a_{\tilde{\vec{k}}}^{t\,\dagger}]
&=(2 \pi)^3 \delta_{s t} \delta^{(3)} (\vec{k}-\tilde{\vec{k}})\, , 
\\
[b_{\vec{k}}^s,b_{\tilde{\vec{k}}}^{t\,\dagger}]
&=-(2 \pi)^3 \delta_{s t} \delta^{(3)} (\vec{k}-\tilde{\vec{k}})\, .
\end{align}
Thus the states created by $b_{\vec{k}}^{\dagger}$ are of negative
 norm, as expected for ghosts.

\section{The wavefunction for a de Sitter background}
\label{sec:The_wavefunction_for_a_de_Sitter_background}

In the case of de Sitter space the definition of the wavefunction in 
the Euclidean path integral formalism 
would normally be done in global coordinates with a foliation by equal 
time slices which are copies of $S^3$. This is the Hartle-Hawking wavefunction \cite{Hartle:83}; however, 
there is a simplification in our case.  We know~\cite{Clunan:2009ib} that for wavelengths 
passing through the horizon sufficiently late,\footnote{That is,
sufficiently late so that they do not see the 
curvature of the universe at horizon exit.} 
it suffices to use the  coordinates of (\ref{metric}) which cover only 
half the spacetime, resulting in a wavefunction similar to, and extending, that in \cite{Maldacena:2002vr}. So, the observables closely approximate those found in the calculation with $S^3$ hypersurfaces, and for all but the $\ell \lesssim 20$ on the $S^2$ of last scattering there is no loss in not doing the calculation in global coordinates. At late times the curvature terms don't affect the background equations of motion and we may use the coordinates with hypersurfaces which are copies of $\mathbb{R}^3$; this is where we use the Euclidean formalism to integrate out over the unobserved variable $Q_{\gamma'}$ before rotating to Lorentzian time. The Wheeler de Witt equation, being based on a Hamiltonian formulation, is Lorentzian. It is generally convenient to give our expressions in a Lorentzian form and point out the changes in the Euclidean form where relevant.

This wavefunction describing fluctuations 
about a de Sitter background may be found by assesing the action on a 
solution of the e.o.m. which has prescribed values of
 $Q_{\gamma},Q_{\gamma'}$ (on a late time surface at $\eta_0$) and
 vanishes in the upper half $\eta$ plane. This last condition restricts 
us to the modes in (\ref{mode_solutions_1}), (\ref{mode_solutions_2}),
 thus the $\gamma_{\vec{k}}^{s}$ in (\ref{fourier}) is
\begin{equation}
\label{f_cpt}
\gamma_{\vec{k}}^{s}=
\frac{\Big(
Q_{\gamma_{\vec{k}}^{s\,'}}(u(\eta_0)v(\eta)- v(\eta_0)u(\eta))
-Q_{\gamma_{\vec{k}}^{s}} (u'(\eta_0)v(\eta)-v'(\eta_0)u(\eta))\Big)}
{u(\eta_0)v'(\eta_0)-v(\eta_0)u'(\eta_0)}.
\end{equation}
From this it is clear that any normalization of the modes will cancel out 
and thus doesn't concern us here. Reality of the field and its time 
derivative at $\eta_0$ require 
$Q_{\gamma_{-\vec{k}}^{s}}=(Q_{\gamma_{\vec{k}}^{s}})^{*}$ and
 $Q_{\gamma_{-\vec{k}}^{s\, '}}=(Q_{\gamma_{\vec{k}}^{s\, '}})^{*}$. 
We are assessing the action on a solution of the eom and hence we are 
dealing with a boundary term on the future boundary:\footnote{It is noted
in \cite{Marinov:80} that the pre-exponential term in the wavefunction in the
quadratic case is independent of the
canonical coodinates; this is borne out by our result.}
\begin{eqnarray}
\imath S|_{\eta_0} 
&=& \imath\!\sum_{s=\pm} \int \frac{d^3 k}{(2 \pi)^3}  {k^3 \over  H^2} 
\Biggl[  -{1\over 4  z^2} \gamma_{\vec{k},z}^s \gamma_{-\vec{k}}^s 
+\beta \left(\gamma_{\vec{k},zzz}^s \gamma_{-\vec{k}}^s-\gamma_{\vec{k},zz}^s
 \gamma_{-\vec{k},z}^s+2 \gamma_{\vec{k}}^s 
\gamma_{-\vec{k},z}^s\right)\Biggr]\Big|_{z_0} 
\label{boundary_action}
\\
&=& \imath\!\sum_{s=\pm} \int \frac{d^3 k}{(2 \pi)^3}  {k^3 \over  H^2}
 \Biggl[
|Q_{\gamma_{\vec{k}}^s}|^2 \bar{A} 
-\frac{1}{2 k} 
\left( Q_{\gamma_{\vec{k}}^{s\, '}}Q_{\gamma_{\vec{k}}^{s}}^*
 + Q_{\gamma_{\vec{k}}^{s\, '}}^* Q_{\gamma_{\vec{k}}^{s}}\right) \bar{B}
+\frac{1}{k^2} |Q_{\gamma_{\vec{k}}^{s\, '}}|^2 \bar{C}
\Biggr]\,,
\end{eqnarray}
where the coefficients,
\begin{align}
\label{abar}
\bar{A}&=\beta\left(\frac{\left(u_{,zzz} v_{,z}-v_{,zzz} u_{,z}\right)}
{u v_{,z}-v u_{,z}}\right)\Big|_{z_0}\,,
\\
\label{bbar}
\bar{B}&=\left(\frac{-1}{4 z^2}+ \beta \frac{u v_{,zzz} -v u_{,zzz}
+ 2 u v_{,z} - 2 v u_{,z} +u_{,z}v_{,zz}- v_{,z}u_{,zz}}{u v_{,z}
 - v u_{,z}}\right)\Big|_{z_0}\,,
\\
\label{cbar}
\bar{C}&=\beta\left(\frac{v u_{,zz}-u v_{,zz}}{u v_{,z}-v u_{,z}}\right)
\Big|_{z_0}\,,
\end{align}
are obtained by the insertion of (\ref{f_cpt}) into (\ref{boundary_action}).
This leaves us with a wavefunction
\begin{eqnarray}
&&\Psi(Q_{\gamma_{\vec{k}}^s},Q_{\gamma_{\vec{k}}^{s\, '}})
= N(\eta_0)
\exp( \imath S_{\vec{k}} |_{\eta_0})
\cr
&&\qquad
=N(\eta_0) \exp\Biggl[ \imath  {k^3 \over  H^2}
 \Biggl(|Q_{\gamma_{\vec{k}}^s}|^2 \bar{A} 
-\frac{1}{2 k}
 \left( Q_{\gamma_{\vec{k}}^{s\, '}}Q_{\gamma_{\vec{k}}^{s}}^*
 + Q_{\gamma_{\vec{k}}^{s\, '}}^* Q_{\gamma_{\vec{k}}^{s}}\right) \bar{B}
+\frac{1}{k^2} |Q_{\gamma_{\vec{k}}^{s\, '}}|^2 \bar{C} \Biggr)
\Biggr]\,.
\label{full_wavefunction}
\end{eqnarray}

The Wheeler-de-Witt equation, constructed from (\ref{classical_hamiltonian}),
 $\mathscr{H} \Psi = \imath \partial_{\eta} \Psi$ and
 $P = - \imath \partial_{Q}$, is then  equivalent to 
\begin{align}
\frac{1}{4 z^2}+\frac{\bar{B}^2}{4 \beta}-\beta = \bar{A}_{,z}\,,
\\
-2\bar{A}+\frac{\bar{B}\bar{C}}{\beta}=\bar{B}_{,z}\,,
\\
-\bar{B}-\frac{1}{4 z^2}+\frac{\bar{C}^2}{\beta}+2\beta=\bar{C}_{,z}\,,
\end{align}
which are satisfied by (\ref{abar}), (\ref{bbar}), (\ref{cbar}).

\subsection{The small $\beta$ approximation}
\label{section:small_beta}

We will need to calculate $\bar{A}, \bar{B}, \bar{C}$ of (\ref{abar}), (\ref{bbar}), (\ref{cbar}), though little useful progress can be made employing the full form of the mode (\ref{normalised_mode_v}). This is facilitated by only considering the physical case where $\beta\approx 0$ (since $H\approx 0$ and one can expect $\alpha$ is of order one).

Once derivatives are neglected we are only dealing with expressions which are homogeneous of degree zero in $u$ and $v$, so we can drop overall factors and use $u, v$ in the form (\ref{mode_solutions_1}), (\ref{mode_solutions_2}). This can be further simplified by using
\begin{equation}
\label{hankel_breakdown}
H_{\lambda}^{(2)} (z) = \frac{J_{-\lambda}(z)
 -e^{\imath \pi \lambda} J_{\lambda}(z)}{-\imath \sin (\lambda \pi)}\,,
\end{equation}
where $\lambda = \sqrt{1+1/\beta}/2$, and dropping the factor of $\imath/\sin(\lambda \pi)$. With real $z$ and imaginary $\lambda$ the complex conjugate of $J_{-\lambda}(z)$ is $J_{\lambda}(z)$. Since $\beta$ is negative the $e^{\imath\pi\lambda}$ suppresses the second term in (\ref{hankel_breakdown}) more than any power of $\beta$. The late time (small $z$) expansion for the Bessel functions
\begin{equation}
J_{\lambda}(z)
=\frac{z^{\lambda}}{2^{\lambda} \Gamma(1+\lambda)}
 \Big(1-\frac{z^2}{4 (1+\lambda)}+\frac{z^4}{32 (1+\lambda) (2+\lambda)} 
+\ldots \Big)\,,
\end{equation}
shows us we can drop another factor, and we are left with the important part of the ghost mode being
\begin{equation}
z^{\frac{3}{2}-\lambda}
 \Big(1-\frac{z^2}{4 (1-\lambda)}+\frac{z^4}{32 (1-\lambda) (2-\lambda)} 
+\ldots \Big)\,,
\end{equation}
The first few terms of the resulting late time, small $\beta$ series are 
\begin{align}
\bar{A} &= -\frac{1}{4 z}+\imath \left(\frac{1}{4}-2 \beta \right)
+\frac{1}{4}\left( 1+2\beta\left(-1+\sqrt{1+\frac{1}{\beta}}\right)\right) z
 + \ldots\,,
 \\
\bar{B} &= -\beta \left( 1+\sqrt{1+\frac{1}{\beta}} \right)
 +\imath \beta \left(3+\sqrt{1+\frac{1}{\beta}}\right) z +\ldots\,,
 \\
\bar{C} &= \frac{ \left(  \sqrt{1+\frac{1}{\beta}} -1\right) \beta}{2 z}
 + \frac{\beta \left(3+\sqrt{1+\frac{1}{\beta}}\right) z}
{\left(-3+\sqrt{1+\frac{1}{\beta}}\right)} +\ldots\,.
\end{align}

\section{Observables from tensor perturbations}
\label{sec:tensor_observables}
\subsection{The usual gravity}

Here we make a few notes about gravity where the Lagrangian consists only of
 the Einstein-Hilbert term.  From equation (\ref{de_sitter_action})
 the second order part of the action is 
\begin{align}
S=&\int d\eta d^3 x {1\over 8} e^{2 \rho} 
\Big(\gamma'_{i j}\gamma'_{i j}-\gamma_{i j,k}\gamma_{i j,k}\Big),
\end{align}
with Fourier components as in equation (\ref{2nd_order_ds_fourier_action})
 given by
\begin{equation}
S_{\vec{k}}=\sum_{s=\pm} \int d \eta {1\over 4} e^{2 \rho}
 (|\gamma_{\vec{k}}^{s\,'}|^2-k^2 |\gamma_{\vec{k}}^{s}|^2)\,.
\end{equation}
In this case, of course, we have canonical coordinates 
$Q_{\gamma_{\vec{k}\,r}^s}=\gamma_{\vec{k}\,r}^s$ and
 $P_{\gamma_{\vec{k}\,r}^s}={1\over 2}e^{2\rho} \gamma_{\vec{k}\,r}^{s\,'}$;
 and the Hamiltonian is given by
\begin{eqnarray}
\mathscr{H}_{\vec{k}}^s 
&=&
{e^{2\rho}\over 4}  |\gamma_{\vec{k}}^{s\,'}|^2 
+{e^{2\rho} k^2 \over 4}  |\gamma_{\vec{k}}^s|^2 
\cr
&=&
e^{-2 \rho} |P_{\gamma_{\vec{k}}^{s}}|^2
+{e^{2\rho} k^2\over 4}  |Q_{\gamma_{\vec{k}}^s}|^2\,.
\end{eqnarray}
The wavefunction is constructed as in 
section (\ref{sec:The_wavefunction_for_a_de_Sitter_background}), where 
equation (\ref{f_cpt}) takes the simplified form,
\begin{equation}
\gamma_{\vec{k}}^{s}=
\frac{Q_{\gamma_{\vec{k}}^{s}} u(\eta)}
{u(\eta_0)},
\end{equation}
and is plugged into a simplified form of equation (\ref{boundary_action}),
\begin{align}
\imath S|_{\eta_0} 
= -\imath\!\sum_{s=\pm} \int \frac{d^3 k}{(2 \pi)^3} 
 {k^3 \over 4  H^2 z^2} \gamma_{\vec{k},z}^s \gamma_{-\vec{k}}^s \,,
\end{align}
to give a wavefunction
\begin{align}
\label{usual_gravity_wavefunction}
\Psi_{\gamma_{\vec{k}}^s} (Q_{\gamma_{\vec{k}}^s}) 
= N \exp \left( \frac{-k^3}{4 H^2} \left( \frac{1}{1+z^2}
 +\frac{\imath}{z(1+z^2)}\right) |Q_{\gamma_{\vec{k}}^s}|^2 \right).
\end{align}
From this we see that the probability distribution for $Q_{\gamma_{\vec{k}}^s}$,
\begin{align}
P(Q_{\gamma_{\vec{k}}^s}) = |\Psi_{\gamma_{\vec{k}}^s}|^2
 = |N|^2 \exp\left( \frac{-k^3}{2 H^2 (1+z^2)} |Q_{\gamma_{\vec{k}}^s}|^2 \right)
\end{align}
freezes out at late times ($z \downarrow 0$); hence, observables
involving $Q_{\gamma_{\vec{k}}^s}$ all freeze out. We note that in order
for the various correlation functions of $\gamma$ to freeze out only the real
component of the argument of the wavefunction (\ref{usual_gravity_wavefunction})
needs to freeze out. This is also observed in \cite{Maldacena:2002vr}.  
We will see that a similar situation arises for the two point function when 
we include the Weyl squared term.\footnote{Linear combinations of the 
classical mode solutions in equations (\ref{mode_solutions_1}),
(\ref{mode_solutions_2}) freeze out since they start with a constant and 
have no term of order one in $z$, thus we must have freeze-out of 
observables involving $\gamma$.}

\subsection{With the Weyl squared term}
The probability distribution the wavefunction of 
equation (\ref{full_wavefunction}) leads to is
\begin{eqnarray}
P(Q_{\gamma_{\vec{k}}^s},Q_{\gamma_{\vec{k}}^{s\, '}})
&=&|N(\eta_0)|^2 \exp\left( \imath S_{\vec{k}}|_{\eta_0} 
+(\imath S_{\vec{k}}|_{\eta_0})^*  \right)
\cr
&=&|N(\eta_0)|^2 \exp\left(\frac{k^3}{H^2} 
\left(|Q_{\gamma_{\vec{k}}^s}|^2 \tilde{A} -\frac{1}{2 k} 
\left( Q_{\gamma_{\vec{k}}^{s\, '}}Q_{\gamma_{\vec{k}}^{s}}^*
 + Q_{\gamma_{\vec{k}}^{s\, '}}^* Q_{\gamma_{\vec{k}}^{s}}\right) \tilde{B}
+\frac{1}{k^2} |Q_{\gamma_{\vec{k}}^{s\, '}}|^2 \tilde{C}
\right)\right)
\cr
&=&|N(\eta_0)|^2 \exp\left(\frac{k^3}{H^2}
 \left( \left(\tilde{A}-\frac{\tilde{B}^2}{4\tilde{C}}\right)
 |Q_{\gamma_{\vec{k}}^s}|^2+\frac{\tilde{C}}{k^2}
\Big|Q_{\gamma_{\vec{k}}^{s\, '}}-\frac{k \tilde{B}}{2\tilde{C}}
Q_{\gamma_{\vec{k}}^s}\Big|^2   \right)\right)\,,
\end{eqnarray}
where $\tilde{A}=2 \Re (\imath \bar{A})$, etc.

As before (section \ref{section:small_beta}) we may do a late time expansion of 
the ghost mode to select the dominant terms when $\beta$ is small and negative,
\begin{align*}
P(Q_{\gamma_{\vec{k}}^s},Q_{\gamma_{\vec{k}}^{s\, '}})
&=|N(\eta_0)|^2 \exp\Big( \frac{k^3}{H^2} \Big(
\Big( -\frac{1}{2}+4\beta
 +\left(\frac{1}{2}-4\beta\right)z^2 +\ldots \Big) |Q_{\gamma_{\vec{k}}^s}|^2
 \\
&\qquad+\frac{1}{k^2} \Big(\imath \sqrt{1+\frac{1}{\beta}}\beta\frac{1}{z} 
+\frac{10\imath \sqrt{1+\frac{1}{\beta}} \beta^2 z}{(-1+3\beta)} 
+ \frac{2\beta (1+10\beta) z^2}{(-1+8\beta)} + \ldots \Big)\times \\
&\qquad\qquad\qquad\qquad\qquad\qquad\qquad\Big|Q_{\gamma_{\vec{k}}^{s\, '}}
+k \Big(z - \frac{3\imath z^2}{\sqrt{1+\frac{1}{\beta}}}
 +\ldots\Big) Q_{\gamma_{\vec{k}}^s}\Big|^2 \Big)
\Big)\,.
\end{align*}
Since $-1<\beta<0$ and $z>0$ it looks like the dependence on $Q_{\gamma'}$ 
is of the wrong sign to be integrated out. This problem does not emerge in the
 Euclidean formalism (as noted in \cite{Hawking:2001yt}). We can rotate into 
the Euclidean and thus get an extra minus sign, which allows us to integrate 
over the unobserved variable. This leaves us with
\begin{equation}
\label{probability}
P(Q_{\gamma})
=|N(\eta_0)|^2 \exp\left[ \frac{k^3}{H^2} 
\left(\frac{-1}{2} - 4 H^2\alpha^2 +\left(\frac{1}{2}+4 H^2 \alpha^2\right)z^2 
+\ldots \right) |Q_{\gamma_{\vec{k}}^s}|^2 \right]\,.
\end{equation}
The simple $k$ dependence here is due to the fact that the $k$ dependence 
can be removed from the eom that the modes satisfy; in the case of the flat 
space wavefunction the $k$ dependence is not an overall factor. So observables 
freeze out at late times, with the two point function having a leading
contribution for large $k$ of 
\begin{equation}
<|\gamma_{\vec{k}}^{s}|^{2}>=\frac{H^2}{k^3} \left(1-8 H^2\alpha^2
  +\ldots\right)\,,
\end{equation}
for the tensor perturbations $\gamma_{\vec{k}}^{s}$. 

\section{Conclusions}
\label{sec:conc}

We have shown that in higher derivative theories of gravity it is possible to live with ghosts, so that there is no need for back-substitution. So the result of taking 
the theory with the Weyl squared term seriously is not that we see corrections 
at order $O(H)$, as we might have expected given the results for a higher 
derivative simple harmonic oscillator; instead, as in the case of 
back-substitution we get corrections at order $O(H^2)$.  Thus, if we only 
consider the addition of a Weyl squared term to the action we see no 
qualitative difference between back-substitution and taking the full theory seriously, as we do here. This similarity may not extend further: further work is required to  determine whether the unusually high cut-off found by back-substitution \cite{breakdown} would be removed in our approach. 

\acknowledgements
This work was supported in part by JSPS Grant-in-Aid for Scientific 
Research (A) No.~21244033, by JSPS Grant-in-Aid for Creative 
Scientific Research No.~19GS0219,
and by Monbukagaku-sho Grant-in-Aid for the global COE program at
Kyoto University,
"The Next Generation of Physics, Spun from Universality and Emergence". TC would like to thank the Yukawa Institute at Kyoto University for their  hospitality while this work was being carried out.


\providecommand{\href}[2]{#2}\begingroup\raggedright\endgroup

\end{document}